\documentclass[12pt]{amsart}
\usepackage{amsmath}
\usepackage{amsthm}
\usepackage{amsfonts}
\usepackage{graphicx}
\usepackage{amssymb} 
\usepackage{dsfont}
\newenvironment{nouppercase}{
  
  \renewcommand{\uppercasenonmath}[1]{}}{}

\begin{document}

\title[On the quantization of some polynomial minimal surfaces]
{On the quantization of some polynomial minimal surfaces}
\author[Jens Hoppe]{Jens Hoppe}
\address{Braunschweig University, Germany}
\email{jens.r.hoppe@gmail.com}

\begin{abstract}
A class of exact membrane solutions is quantized.
\end{abstract}

\begin{nouppercase}
\maketitle
\end{nouppercase}
\thispagestyle{empty}
\noindent 
After half a century of unsuccessful searches for exact compact 2-dimensional membrane solutions sweeping out an algebraic 3 dimensional manifold,
\begin{equation}\label{eq1} 
u(x^{\mu}) := t^2-x^2-y^2-z^2-c(t+z)^6 = 0
\end{equation}
was recently shown to define zero--mean--curvature hypersurfaces in $\mathbb{R}^{1,3}$ \cite{1}. In this note I would like to point out possible quantizations of (\ref{eq1}). First let us slightly generalize a known procedure (cp. \cite{2},\cite{3}) for axially symmetric surfaces: defining
\begin{equation}\label{eq2} 
\lbrace F(t,\vec{x}), G(t,\vec{x})\rbrace := -\frac{1}{2}\vec{\nabla}u \cdot \vec{\nabla}F \times \vec{\nabla}G 
\end{equation}
one finds
\begin{equation}\label{eq3} 
\begin{split}
\lbrace x,y \rbrace & = -\frac{1}{2}(2t-f-\kappa f') = -t+\kappa+3\kappa^5c\\
\lbrace y,\kappa \rbrace & = x \qquad \lbrace \kappa, x \rbrace = y
\end{split}
\end{equation}
where $f(x) := \kappa + c\kappa^5$, $\kappa := z+t$;\\
which naturally leads to the problem of finding hermitean $N \times N$ matrices $X, Y$ and $\Lambda$, resp. $W = X+iY$, $W^{\dagger} = X-iY$ and $\Lambda$, satisfying
\begin{equation}\label{eq4} 
\begin{split}
[\Lambda, W] & = \hbar W, \quad [W, W^{\dagger}] = -\hbar D(\Lambda) \\
D(\Lambda) & := 2t-f(\Lambda) - \Lambda f'(\Lambda) = 2(t-\Lambda-3c\Lambda^5). 
\end{split}
\end{equation}
With $\Lambda$ diagonal and $W$ non--zero only on the first lower off--diagonal $(W_{i+1,i} = w_i \; i=1,2,\ldots ,N-1)$, one finds
\begin{equation}\label{eq5} 
\begin{split}
\lambda_i & = \lambda_1 + (i-1)\hbar, \quad i= 1,2,\ldots, N \\
\hbar d_i & = 2 \hbar(t-\lambda_i-3c\lambda_i^5) = (r_i^2 - r_{i-1}^2),\; r_0 \equiv 0 \equiv r_N,
\end{split}
\end{equation} 
hence 
\begin{equation}\label{eq6} 
r^2_k := |w_k|^2 = \hbar \sum_{j=1}^k d_j \quad k = 1,2, \ldots, N
\end{equation}
$\text{Tr}D = \sum_1^N d_j = 0$ follows both from (\ref{eq6}), as $w_N \equiv 0$, and from (\ref{eq4}) (the trace of a commutator of finite matrices having to be 0). So
\begin{equation}\label{eq7} 
Nt = \sum \lambda_i + 3c \sum_i \lambda_i^5
\end{equation}
Choosing $c>0$, $\kappa \in [0, \kappa_M(t)]$ for $t>0$, $f(\kappa_M):= 2t$, $\lambda_1 \rightarrow 0$ as $N \rightarrow \infty$ and $(N-1)\hbar + \lambda_1 \approx \kappa_M - \lambda_1$, resp., when inserting (\ref{eq5}) into (\ref{eq7}), 
\begin{equation}\label{eq8} 
2t \approx (N-1)\hbar\big( 1+ c(\hbar N)^4 (1-\frac{1}{N})(1-\frac{1}{N}-\frac{1}{2N^2}) \big).
\end{equation}
As classically the surface is parametrized by 
\begin{equation}\label{eq9} 
z = -t + \kappa, \qquad r^2 = \kappa(2t -f(\kappa))
\end{equation}
(where $\kappa \in [0, \kappa_M]$ for $t > 0$, and $[-\kappa_M, 0]$ if $t < 0$) the above nicely matches the classical situation: the $d_i = \frac{r_i^2-r_{i-1}^2}{\hbar}$ $\;(\hat{=} 2rr')$ are first positive, then negative, $r_i^2 \big(= \frac{1}{2}(WW^{\dagger} + W^{\dagger}W)_{ii} = (X^2+Y^2)_{ii}\big)$ being the square of the radius of the spherical surface at height $-t + (\lambda_1 + (i-1)\hbar)$.\\
The only difficult part is to obtain $a:=\lambda_1$ and $\hbar$ explicitly as functions of $N$ (and $t$). The question of Quantum--Casimir can be approached in various ways: using (\ref{eq4}),
\begin{equation}\label{eq10} 
\begin{split}
[\frac{1}{2}(WW^{\dagger} + W^{\dagger}W)  - \Lambda(2t - & f(\Lambda)),W]  =: [C_0,W]\\
 & = c \big( [\Lambda^6, W] - 3\hbar(W\Lambda^5 + \Lambda^5 W)\big)\\
 & = \hbar^2 c \big( 2[W,\Lambda^4] + \Lambda(W\Lambda^2-\Lambda^2 W)\Lambda\big) 
\end{split}
\end{equation}
Hence 
\begin{equation}\label{eq11}
\begin{split} 
[C_0 + 2\hbar^2 c \Lambda^4, W] & = -\hbar^3 c \Lambda (\Lambda W + W \Lambda)\Lambda \\
& = -\hbar^2 c [Q_4, W]
\end{split}
\end{equation}
Using
\begin{equation}\label{eq12} 
\Lambda W = W(\Lambda + \hbar), \qquad W\Lambda = (\Lambda-\hbar)W
\end{equation}
the Ansatz $Q_4(\Lambda) = \tilde{Q}_4 (\tilde{\Lambda} := \frac{\Lambda}{\hbar} = x) = \sum_{j=0}^4 \gamma_jx^j$ together with
\begin{equation}\label{eq13} 
WF(x) = F(x-1)W
\end{equation}
easily yields $Q_4(\Lambda) = \frac{1}{2}(\Lambda^4 - \hbar^2 \Lambda^2)$, hence
\begin{equation}\label{eq14} 
[Q := C_0 + \frac{5c}{2}\hbar^2\Lambda^4 - \frac{c}{2}\hbar^4\Lambda^2, W] = 0.
\end{equation}
To doublecheck, and calculate the value $q$  ($Q= q\mathds{1}$), one calculates (from (6))    
\begin{equation}\label{eq15}
\begin{split} 
-r^2_{k+1} & = -(W^{\dagger} W)_{k+1} = 2\hbar \sum_{j=1}^{k+1}(\lambda_j+3c\lambda_j^5-t)\\
 & = 2\hbar \big( (k+1)(a-t+3ca^5) + \hbar \frac{k(k+1)}{2} \\
 & \quad \qquad  +3c \big(  5a^4\hbar\frac{k(k+1)}{2} + \frac{5}{3}a^3\hbar^2(2k^3+3k^2+k) \\
 & \qquad \qquad \quad \; +\frac{5}{2}a^2\hbar^3(k^4+2k^3+k^2) \\
 & \qquad \qquad \quad \; + \frac{1}{6}a\hbar^4(6k^5+15k^4+10k^3-k)\\
 & \qquad \qquad \quad \; + \frac{\hbar^5}{12} (2k^6 + 6k^5 + 5k^4 -k^2)   \big) \big)
\end{split}
\end{equation}
and writes the Quantum Casimir as
\begin{equation}\label{eq16}
\begin{split} 
Q & = W^{\dagger} W - \hbar^2 \tilde{g}( \frac{\Lambda}{\hbar}) (= C_0 + \frac{5c}{2}\hbar^2\Lambda^4-\frac{\hbar^4}{2}\Lambda^2c)\\
-\tilde{g}(x) & := -\frac{t}{\hbar}+x + x^2 -2\frac{tx}{\hbar}+c\hbar^4(x^6+3x^5+\frac{5}{2}x^4-\frac{1}{2}x^2) 
\end{split}
\end{equation}
Inserting (\ref{eq15}), and $\frac{\Lambda}{\hbar} = \frac{a}{\hbar}\mathds{1} + 
\left(\begin{smallmatrix}
0 &  &  &  & \\
  &1 &  &  & \\
  &  &2 &  & \\
  &  &  & \ddots  & \\
  &  & &  & N-1  \\
\end{smallmatrix}\right)$
one can check that indeed all powers $k^{i = 1, \ldots, 6}$ (i.e. all $k$--dependent terms cancel), hence $Q = q \cdot \mathds{1}$, provided 
(from $O(k^0)$)
\begin{equation}\label{eq17}
\begin{split} 
q & = (a\hbar + a^2 - 2at) - 2a\hbar + 2\hbar t - t\hbar \\
  & \quad  + c (a^6 - 3a^5\hbar + \frac{5}{2}a^4\hbar^2 - \frac{a^2}{2}\hbar^4), 
\end{split}
\end{equation}
\\
\begin{equation}\label{eq18}
\begin{split} 
q & = (a\hbar + a^2 -2at)+\hbar^2(N-1)(N-2)+2\hbar(N-1)(a-t)\\
  & \quad + c\hbar^6 \big( (\frac{a}{\hbar}+N-1)^6 + 3(\frac{a}{\hbar}+N-1)^5 + \frac{5}{2}(\frac{a}{\hbar}+N-1)^4 \\
  & \qquad \qquad \quad- \frac{1}{2}(\frac{a}{\hbar}+N-1)^2 \big) - t\hbar
\end{split}
\end{equation}
(from $-\tilde{g}(\lambda_N)\hbar^2 = q$), and
\begin{equation}\label{eq19}
\begin{split} 
 t & = a + (N-1)\frac{\hbar}{2} \\
   & \quad + 3c \big( a^5 + \frac{5}{2}a^4\hbar(N-1) + \frac{5}{3}a^3\hbar^2(N-1)(2N-1)\\
   & \qquad \qquad \;  +\frac{5}{2}a^2\hbar^3N(N-1)^2 + \frac{1}{6}a\hbar^4(N-1)(2N-1)(3N^2-3N-1)\\
   & \qquad \qquad \;  +\frac{\hbar^5}{12}N(N-1)^2(2N^2-2N-1) \big)
\end{split}
\end{equation}
(from (\ref{eq7})).
One can check that these 3 equations admit solutions $a$, $\hbar$ and $q$ going to zero as $N \rightarrow \infty$, hence determining solutions of (\ref{eq4}) that are non--commutative analogues of time--dependent spherical surfaces sweeping out zero--mean curvature hypersurfaces in $\mathbb{R}^{1,3}$. 
(Note that while the power 6 in (\ref{eq1}) is needed for the mean curvature to vanish, the quantization procedure, i.e, the way to obtain exact solutions of (\ref{eq4}) would work in many different situations; including any $f$, cp. (\ref{eq3})$_{c>0}$, that is monotonic). \\
Light--cone quantization (i.e. foliating with respect to $\tau := \frac{t+z}{2}$) on the other hand gives $X_1 = \gamma \tau Q$,  $X_2 = \gamma \tau P$, $[Q, P] = i\hbar$, $X_- = \gamma^2 \tau \big(\frac{1}{2}(Q^2 + P^2) + \gamma^2 \frac{\tau^4}{10}\big)$ (quantizing the velocity potential of spatially homogeneous $2+1$ dimensional Chaplygin--Karman--Tsien fluids), $2X_- \tau - X_1^2 - X_2^2 = \gamma^4 \frac{\tau^6}{5}\mathds{1}$ (without any $\hbar$--corrections), and 
\begin{equation}\label{eq20}
\Delta X_- := \ddot{X}_- + \sum^2_{i=1} \big[ [X_-, X_i], X_i \big]\frac{1}{\hbar^2} = 0
\end{equation}  
(just as $\Delta X_i = 0$), which is very interesting, and potentially signalling a flaw in common beliefs concerning lack of Lorentz--invariance in non--commutative membrane theory.

\vspace{1cm}
\noindent
\textbf{Acknowledgement.} I am grateful to J.Arnlind for very helpful discussions.

\end{document}